\documentclass[useAMS,usenatbib]{mn2e}
\usepackage{graphicx}
\title[Gl\,86\,B: a white dwarf orbits an exoplanet host star]{Gl\,86\,B: a white dwarf orbits an exoplanet host star}
\author[M. Mugrauer and R. Neuh\"auser]{M. Mugrauer$^{1}$\thanks{E-mail: markus@astro.uni-jena.de (MUG)
and rne@astro.uni-jena.de (NEU)} and R. Neuh\"auser$^{1}$\footnotemark[1]
\thanks{Based on observations obtained at the European Southern Observatory on Paranal in ESO programs
074.C-0262(A) and 070.C-0179(A), the latter taken from the public archive.}\\
$^{1}$Astrophysikalisches Institut, Universit\"at Jena, Schillerg\"a{\ss}chen 2-3, 07745 Jena,
Germany\\}
\begin{document}

\date{Accepted , Received }

\pagerange{\pageref{firstpage}--\pageref{lastpage}} \pubyear{2005}

\maketitle

\label{firstpage}

\begin{abstract}
In this letter we present our first high contrast NACO/SDI observations of the exoplanet host star
Gl\,86 and results from NACO spectroscopy. Els et al. (2001) found a faint co-moving companion
located only $\sim$\,2\,arcsec east of the exoplanet host star Gl\,86\,A. Our high contrast SDI
observations rule out additional stellar companions from 1\,AU up to 23\,AU, and are sensitive for
faint T dwarf companions down to 35\,M$_{Jup}$. We present evidence for orbital motion of Gl\,86\,B
around the exoplanet host star Gl\,86\,A, which finally confirms that this is a bound binary
system. With the given photometry from Els et al. (2001) and the obtained NACO spectroscopy we
prove that the companion Gl\,86\,B is a cool white dwarf with an effective temperature of
5000$\pm$500\,K. This is the first confirmed white dwarf companion to an exoplanet host star and
the first observational confirmation that planets survive the post main sequence evolution of a
star from which they are separated by only one to two dozen AU (giant phase and planetary nebula)
as expected from theory.
\end{abstract}


\section{Introduction}

Gliese\,86 (thereafter Gl\,86) is a K1 dwarf located at a distance of 10.9$\pm$0.08\,pc (Hipparcos,
ESA 1997). Queloz et al. (2000) found a 15.8\,day periodical variation of its radial velocity.
Because Gl\,86 does not show any chromospheric activity or photometric variability they concluded
that the variation of the radial velocity is induced by an exoplanet with a minimum mass of
4\,$M\rm_{Jup}$ on an almost circular orbit (e=0.046, a=0.11\,AU). Furthermore they reported a
long-term linear trend in the radial velocity data (0.5 m s$^{-1}$ d$^{-1}$) observed over a time
span of 20 years in the CORAVEL program and also confirmed by CORALIE measurements (0.36 m s$^{-1}$
d$^{-1}$). This is a clear signature of a further stellar companion in the Gl\,86 system with a
period longer than 100\,yr (a$\gid$20\,AU).

Els et al. (2001) found a faint companion 2\,arcsec east of Gl\,86 which clearly shares common
proper motion. They also obtained near infrared photometry (J=14.7$\pm$0.2, H=14.4$\pm$0.2, and
K=13.7$\pm$0.2) and from the derived color (J-K$\sim$1) they concluded that Gl\,86\,B must be
substellar with a spectral type between late L to early T. However a substellar companion
(m$\le$78\,$M\rm_{Jup}$) cannot explain the detected long term trend in the radial velocity of the
exoplanet host star.

At the begin of 2005 we started a search for faint substellar companions of all exoplanet host
stars known to harbor several substellar companions, i.e. several planets or as in the case of
Gl\,86 one planet and a reported brown dwarf companion (see Els et al. 2001). Our goal is to detect
additional companions. We carry out our observations with NACO/VLT and its new simultaneous
differential imaging device SDI which is particular sensitive for faint cool substellar companions
exhibiting strong methane absorption features (T dwarfs), yielding a much higher contrast than
standard AO imaging with NACO alone. All our targets are main-sequence stars with ages in the range
of 1 to 10\,Gyrs. From theoretical models (Baraffe et al. 2003) we expected that most of the
substellar companions of our targets are T dwarfs cooler 1400\,K, which are detectable with
NACO/SDI close to the primary stars.

In section 2 we describe our SDI observations, the obtained astrometry and photometry in detail.
With the achieved high contrast SDI imaging we furthermore proof that their is no further companion
in the system which could induce the reported long term trend in the radial velocity of Gl\,86\,A
(Queloz et al. 2000). Spectra of Gl\,86\,B were taken with NACO and were retrieved from the ESO
public archive, yielding the surprising result that this companion is a white dwarf companion. We
presented the spectroscopy in section 3. Finally section 4 summarizes and discusses all results of
this letter.

\begin{figure}\resizebox{\hsize}{!}{\includegraphics{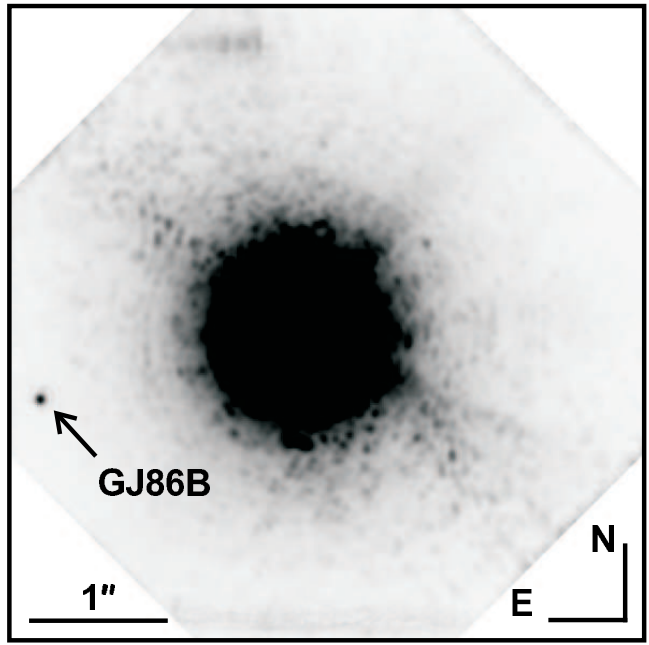}}
\caption{40 min of H band imaging with SDI at the ESO VLT (north up east to the left). This image
is taken through the SDI narrow band filter with a central wavelength of 1.575 $\mu$m. It is one
out of four simultaneously taken images. The bright star in the center is the exoplanet host star
Gl\,86\,A. The about 10\,mag fainter companion Gl\,86\,B is located $\sim$2\,arcsec east of the
primary. Several bright speckles are clearly visible close to the primary star.} \label{sdia}
\end{figure}

\section{NACO/SDI Observations}

Detection of faint objects close to a much brighter source is the main challenge in the direct
imaging search for substellar companions (brown dwarfs or planets) of stars. Within 1\,arcsec of
the bright central source the field is filled with speckles which are residuals from the none
perfect adaptive optics correction of the incoming disturbed wavefront. The achievable signal to
noise in this speckle-noise limited region does not increases with integration time, hence only a
subtraction of the speckle pattern can improve the detection limit close to a bright source.

At the ESO/VLT the simultaneous differential imager (SDI) is offered for the AO-system NACO (Lenzen
et al. 2004). Therein a double Wollaston prism splits the beam, coming from the AO system, into 4
beams which pass then through three different narrow band filters with central wavelengths 1.575,
1.600 and 1.625\,$\mu$m and a bandwidth of 25\,nm bandwidth each. The resulting speckle pattern of
the four images is almost identical. Since cool (T$\le$1400\,K) objects exhibit a strong methane
absorption band at 1.62\,$\mu$m they appear much fainter in the 1.625$\mu$m filter than in the
1.575$\mu$m filter while the bright star and therefore also the speckle pattern has roughly equal
brightness in all images. Subtracting the 1.625\,$\mu$m SDI image from the image taken through the
1.575\,$\mu$m SDI filter will effectively cancel out the speckle pattern of the star while the
signal from the cool companion remains (see e.g. Biller et al. 2004).

We observed Gl\,86 on 12. Jan 2005 with SDI (see Fig.\,\ref{sdia}). Due to the small SDI field of
view (5x5\,arcsec, tilted by 45\,$^{\circ}$), jittering cannot be used for background substraction.
However we apply a small jitter with a box width of only 0.2\,arcsec (10 SDI pixel) to correct for
bad pixels. Per jitter cycle 5 object frames are taken, each is the average of 60 exposures of 2\,s
each. We always adjust the individual integration time so that only the central 9 SDI pixel of the
primary point spread function are saturated. This improves the detection limit for faint companions
at larger separations to the primary. At the end of each jitter cycle a sky-frame (10\,arcsec
offset from the target in Ra and Dec) is taken which is then subtracted from the 5 object frames to
cancel out the bright infrared sky background. The sky-frames are taken in the same way as the the
target-frames, i.e. 60 times 2\,s integrations. The jitter cycle is repeated 4 times which yields a
total on source integration time of 40\,min. To distinguish between faint companions and any
residual speckles we observe each target at the detector position angles 0 and 33\,$^{\circ}$, i.e.
2x40\,min integration time in total.

\begin{figure}\resizebox{\hsize}{!}{\includegraphics{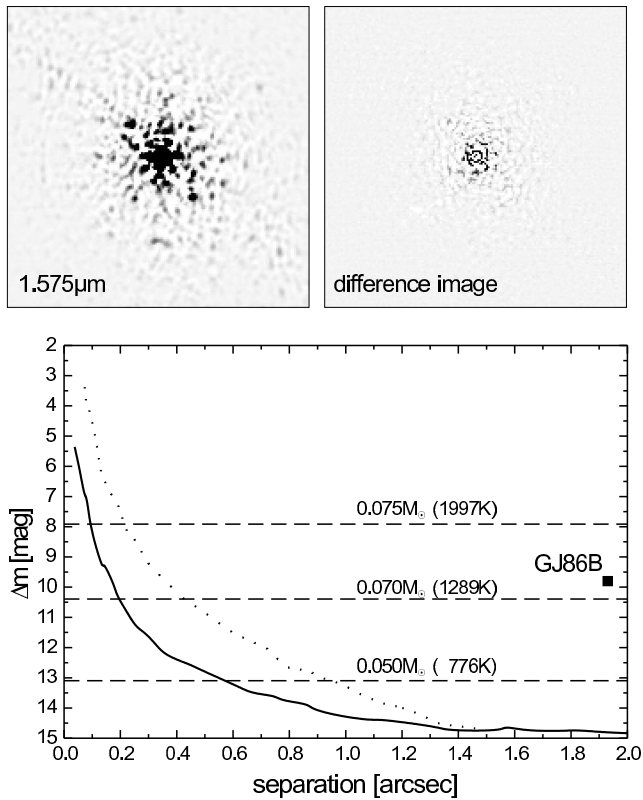}}
\caption{The upper left pattern shows the central part of Fig.\,\ref{sdia} (2.8\,arcsec box width).
The detection limit for this image is illustrated as a dotted line in the bottom panel. The right
pattern shows the subtraction of the difference images (1.575-1.625\,$\mu$m) taken at detector
position angle 0 and 33\,$^{\circ}$. The speckle pattern is effectively subtracted and the achieved
detection limit is improved close to the primary (see solid line). A magnitude difference of
12.8\,mag is reached at 0.5\,arcsec, which is 5.5\,AU in projected separation. The dashed
horizontal lines illustrate the expected magnitude differences for substellar companions derived
with Baraffe et al. (2003) COND models, assuming a system age of 10\,Gyrs. All stellar companions
can be detected beyond 0.1\,arcsec, i.e. 1\,AU in projected separation and T dwarfs come into range
beyond 0.2\,arcsecs (2\,AU). We are sensitive for substellar companions down to
35\,$M\rm{_{Jup}}$}. Gl\,86\,B is well detected in our SDI images and is shown as a black square.
\label{sdib}
\end{figure}

For infrared data-reduction we use the ESO \textsl{Eclipse} package. After flatfielding, we extract
the four SDI quadrants (left-top 1.6\,$\mu$m, right-top 1.575\,$\mu$m, bottom both quadrants
1.625\,$\mu$m) and apply image-registration, shifting, and final averaging on all individual
frames. Because the radial position of the speckles is proportional to the wavelength all SDI
images must be spatially rescaled. Finally the images are aligned, and their flux is adjusted to
eliminate any differences in the quantum efficiency. The 1.575\,$\mu$m and the 1.625\,$\mu$m images
(top-right and bottom-left quadrant) follow the same path through the SDI instrument, i.e. they
provide very similar speckle pattern and yield the best contrast for methane rich companions. We
therefore subtract these two SDI images. The difference frame taken at position angle 0$^{\circ}$
is then subtracted from the difference frame taken at position angle 33\,$^{\circ}$. To filter out
low spatial frequencies all frames are unsharp masked (see Fig.\,\ref{sdib}). The resulting
difference frame compared to the image taken at 1.575\,$\mu$m is shown in Fig.\,\ref{sdib}. The
speckle pattern is effectively subtracted and a detection limit of 12.8\,mag is reached at a
separation of 0.5\,arcsec.

For astrometrical calibration of the SDI camera we observed the binary HIP\,9487. This system is
listed in the Hipparcos catalogue ($\rho$=1.876$\pm$0.001\,arcsec and
$\theta$=278.500$\pm$0.001\,$^{\circ}$ at epoch 1991.25) and accurate astrometry is available for
both components. Therefore the binary separation and position angle can be computed for the given
observing epoch (12 Jan. 2005). We derive a pixelscale of 17.210$\pm$0.087\,mas per pixel. The true
north is slightly rotated to the east by 0.33$\pm$0.24\,$^{\circ}$.

\begin{figure}\resizebox{\hsize}{!}{\includegraphics{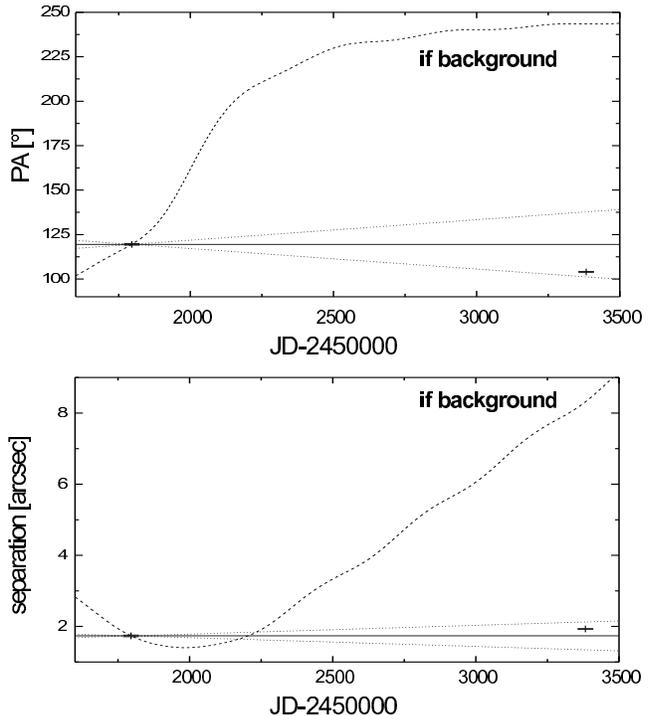}}
\caption{This plot illustrates the astrometric results for the Gl\,86\,A/B system. The first data
point is taken from Els et al. (2001) for epoch 2000. Our SDI astrometry significantly confirms the
common proper motion of Gl\,86\,B to its primary. The dashed lines illustrate the expected change
in position angle and separation if only Gl\,86\,A is moving and Gl\,86\,B is a none-moving
background source. The dotted lines show the expected orbital motion.} \label{astro}
\end{figure}

Due to the large proper and parallactic motion of Gl\,86
($\mu_{\alpha}cos(\delta)$\,=\,2092.59$\pm$0.56\,mas/yr,
$\mu_{\delta}$\,=\,654.49$\pm$0.49\,mas/yr, and $\pi$\,=\,91.63$\pm$0.61\,mas, Hipparcos ESA 1997)
Els et al. (2001) already proved that Gl\,86\,A and B form a common proper motion pair, a result
which is significantly (263$\sigma$ in separation and 279$\sigma$ in position angle) confirmed with
our SDI observations (see Fig.\,\ref{astro}). Gl\,86\,B is located 1.93$\pm$0.01\,arcsec at a
position angle of 104.0$\pm$0.3\,$^{\circ}$. Fig.\,\ref{astro} illustrates the astrometry from Els
et al. (2001) and our astrometric results. The expected change of separation and position angle is
calculated assuming that only Gl\,86\,A is moving and Gl\,86\,B is an unrelated background star
(see dashed lines in Fig.\,\ref{astro}).

By comparing our astrometry with data from Els et al. (2001) we find a significant change in
position angle of $-15.5\pm$0.5\,$^{\circ}$ and +0.196$\pm$0.024\,arcsec in separation for the
given epoch difference. This is a clear evidence for orbital motion. The expected orbital motion
for a companion at a separation of 21\,AU is shown with dotted lines in Fig.\,\ref{astro}
(0.090\,arcsec/yr and 4.2\,$^{\circ}$/yr, see also section\,4).

With the J,H and K infrared magnitudes of Gl\,86\,B given by Els et al. (2001) and the Hipparcos
parallax of Gl\,86\,A we derive the absolute magnitudes of Gl\,86\,B:
$M\rm{_{J}}$=14.5$\pm$0.2\,mag, $M\rm{_{H}}$=14.2$\pm$0.2\,mag, $M\rm{_{K}}$=13.5$\pm$0.2\,mag. If
we assume that Gl\,86\,B is a brown dwarf companion these absolute magnitudes are consistent with a
spectral type L7 to T5 (see Vrba et al. 2004). In our SDI images Gl\,86\,B is comparable bright in
all three filter. We measure flux-ratios $F\rm{_{1.575\,\mu m}}/F\rm{_{1.625\,\mu
m}}$\,=\,1.11$\pm$0.04. This clearly rules out spectral types later than T3, because for theses
spectral types $F\rm{_{1.575\,\mu m}}/F\rm{_{1.625\,\mu m}}>1.3$ due to methane absorption.

\section{NACO Spectroscopy}

\begin{figure}\resizebox{\hsize}{!}{\includegraphics{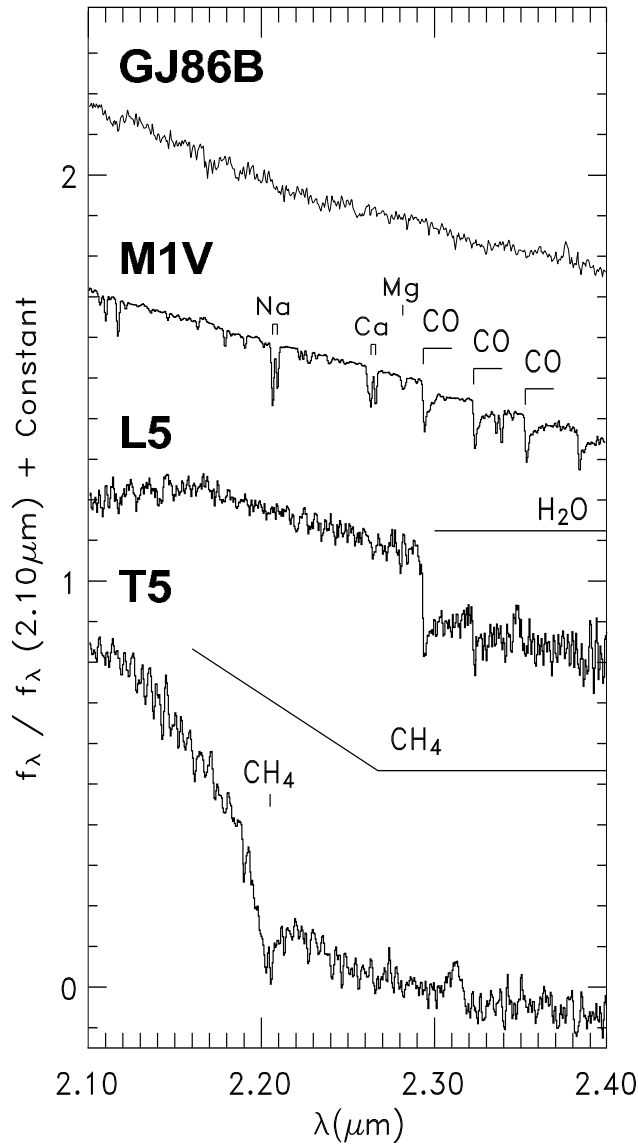}}
\caption{The NACO K band spectrum of Gl\,86\,B with comparison M1V, L5, and T5 comparison spectra
of cool dwarfs from Cushing et al. (2005).} \label{mltspectra}
\end{figure}

Infrared spectra of Gl\,86\,B were taken in ESO observing program 070.C-0173(A) (extracted by us
from the public archive). 8 spectra (120\,s each) were taken in spectroscopic mode S27-SK-3 using
the 86\,mas slit, which yields a resolving power $\lambda/\Delta \lambda$\,=\,1400. For background
subtraction, 8\,arcsec nodding was applied along the slit. To avoid that the bright primary is
located on or close to the slit (saturation) the slit was orientated perpendicular to the position
vector of Gl\,86\,B relative to the primary. All images are flatfielded (lamp flats) and the
spectra are extracted, wavelength calibrated (with Argon lines) and finally averaged, with standard
routines in IRAF. The resulting spectrum is flux calibrated with the photometric standard
HIP\,020677 (G2V). Figure\,\ref{mltspectra} shows the flux calibrated NACO spectrum of Gl\,86\,B,
together with spectra of M1, L5, and T5 dwarfs from Cushing et al. (2005). For the given spectral
range the achieved signal to noise is $\sim$40. Neither characteristic molecular nor atomic
absorption features are visible and the continuum is clearly different to those of L and T dwarfs.
The spectrum is even steeper than the M1V reference spectrum which points to an effective
temperature hotter than 3700\,K. However, in the K-band, the gradient of the continuum is only
slightly varying for effective temperatures higher than 4000\,K, so that we can only give a low
temperature limit for Gl\,86\,B.

\begin{figure}\resizebox{\hsize}{!}{\includegraphics{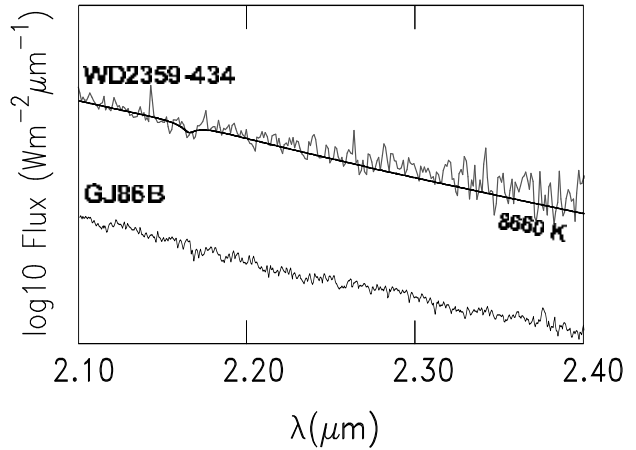}}
\caption{The NACO K band spectrum of Gl\,86\,B with a comparison white dwarf spectra from Dobbie et
al. (2005).} \label{wdspectra}
\end{figure}

Therefore we conclude that Gl\,86\,B is a cool white dwarf companion (see Fig.\ref{wdspectra}). Due
to its high surface gravity ($log(g)>$7) all absorption line are strongly broadened and therefore
hardly detectable (see e.g. Br\,$\gamma$ absorption line at 2.17\,$\mu$m, in Dobbie et al. 2005).
We compare the absolute infrared photometry of Gl\,86\,B with data from Bergeron et al. (2000), who
carried out a detailed photometric and spectroscopic analysis of cool white dwarfs, and derive an
effective temperature of 5000$\pm$500\,K.

\section{Conclusions}

We confirmed common proper motion of Gl\,86\,B to its primary and detected its orbital motion. This
is a clear evidence that Gl\,86\,B is a bound companion of the exoplanet host star with a projected
separation of 21\,AU (1.93$\pm$0.01\,arcsec). With the achieved high contrast SDI detection limit
we can rule out any further stellar companions beyond 0.1\,arcsec up to 2.1\,arcsec, i.e. 1\,AU to
23\,AU in projected separation. T dwarf companions (T$<$1400\,K) can be detected beyond
0.2.\,arcsecs (2\,AU) and we are sensitive for faint substellar companions down to 35\,$M_{Jup}$.
The NACO spectrum of Gl\,86\,B is clearly different to the expected spectral type between L5 and T5
derived from infrared photometry. Gl\,86\,B is faint but its spectrum implies that it is even
hotter than 3700\,K (M1). Therefore we conclude that Gl\,86\,B is a cool white dwarf companion.

Queloz et al. (2000) report a long-term trend in the radial-velocity data of Gl\,86\,A
(0.5\,m\,s$^{-1}$\,d$^{-1}$). This is a clear hint on a further stellar companion in the Gl\,86
system. We compute that with a separation of 21\,AU, the mass of Gl\,86\,B must be
0.55\,$M\rm{_{\sun}}$ to induces this trend in the radial velocity. The derived mass is well
consistent with a white dwarf companion. Santos et al. (2004) determined the mass of Gl\,86\,A to
be 0.7\,$M\rm{_{\sun}}$, hence the total mass of the system is 1.25\,$M\rm{_{\sun}}$ and with a
binary semi-major axis of 21\,AU, the expected orbital time is 86\,yr. This yields an orbital
motion of 0.090\,arcsec/yr and 4.2\,$^{\circ}$/yr, assuming a circular orbit, which is also
consistent with the measured orbital motion (see Fig.\,\ref{astro}).

If we assume that both components of the binary have the same age, the white dwarf progenitor must
be more massive than Gl\,86\,A (0.7\,$M\rm{_{\sun}}$) to be observable today as a white dwarf
companion. According to Weidemann (2000) a 0.55\,$M\rm{_{\sun}}$ white dwarf is the remnant of
1\,$M\rm{_{\sun}}$ star. With the white dwarf models presented by Richer et al. (2000) and the
derived effective temperature of Gl\,86\,B (5000$\pm$500\,K) we can approximate a cooling timescale
between 3 and 6\,Gyr, i.e. the binary system should be 13 to 16\,Gyrs old. For more massive white
dwarf progenitors (2-4\,$M\rm{_{\sun}}$) the system age ranges between 2 to 8\,Gyr.

Gl\,86\,B is the first confirmed white dwarf companion to an exoplanet host star. Theoretically,
planets may or may not survive the red giant and asymptotic giant branch phases of stellar
evolution. Planets which are located outside the red giant's envelope, which reaches about a few
hundred solar radii will survive. Closer companions will be either destroyed or migrate inward and
become a close companion to the white dwarf remnant. According to Burleigh et al. (2002) it seems
likely that distant planets (a$>$5\,AU) survive the late stages of stellar evolution of
main-sequence stars with masses in the range between 1 and 8\,$M_{\sun}$, i.e. all white dwarf
progenitors. In particular in the Gl\,86 system the separation between the white dwarf and the
exoplanet (21\,AU) is large enough that it seems very well possible that the planet can survive the
post main sequence phase of a F or G dwarf.

Furthermore we should mention that Gl\,86 is one of the closest binaries known today to harbor an
exoplanet. Only two other systems $\gamma$\,Ceph (a$\sim$19\,AU, e$\sim$0.36, see Hatzes et al.
2003) and HD\,41004 (23\,AU see Zucker et al. 2004) have comparable separations. In such close
binary systems the dynamical stability of planets is limited to a small region around the planet
host star. According to Holman \& Wiegert (1999) the critical semi-major axis a$\rm{_c}$ for
planets around Gl\,86\,A is only 6.2\,AU assuming a circular binary orbit
(m\,=\,0.55\,$M\rm{_{\sun}}$ and a\,=\,21\,AU). Due to mass loss during the post main sequence
phase of the white dwarf progenitor, the binary separation was even smaller before
(M$\rm{_{tot}}$a\,=\,const $\rightarrow$ a${_{old}}$\,=\,15.4\,AU), i.e. a$\rm{_c}$\,=\,3.7\,AU. We
calculate critical semi-major axis also for the two other close binaries and get
a$\rm{_c}$\,=\,7.5\,AU for HD\,41004 and a$\rm{_c}$\,=\,4.0\,AU for $\gamma$\,Ceph. We should
mention that all exoplanets detected in these close binary systems actually reside within the
proposed long-time stable regions. However it would be of particular interest to search for further
substellar companions in theses close binaries to verify with observational results the published
theoretical constraints of planet stability in binary systems.

\label{lastpage}

\end{document}